# Decoupling heat and electricity: A thermal invisible gateway

Jiahao Li [1,#], Fei Sun [1,#,*], Yichao Liu[1,*], Yawen Qi[1], Qin Liao[1], Jianpu Yang[1], Zhiru Xie[1]

[1]*Key Lab of Advanced Transducers and Intelligent Control System, Ministry of Education and Shanxi Province, College of Physics and Optoelectronics, Taiyuan University of Technology, Taiyuan, 030024 China*

[#] *These authors contribute equally to this work*

[*] Corresponding author: sunfei@tyut.edu.cn

liuyichao@tyut.edu.cn

**ABSTRACT**

The Wiedemann-Franz law dictates an intrinsic coupling between the electrical and thermal conductivities of natural materials, rendering the synergistic realization of high electrical conductivity and superior thermal insulation a longstanding unsolved core challenge in materials science and severely hindering technological advancements in cutting-edge fields such as high-density electronic integration and flexible wearable devices. To address this challenge, we design a thermal invisible gateway based on active thermal metasurfaces (ATMSs) in accordance with the law of energy conservation, achieving precise physical decoupling of thermal and electrical transport pathways. Fabricated on a copper substrate with a symmetric dumbbell-shaped bridge configuration, this structure realizes equivalent suppression of thermal conduction while maintaining unimpeded electrical conduction via the directional compensation and cancellation of heat flux by ATMSs. Both numerical simulations and experiments conducted at room temperature demonstrate that the as-fabricated thermal invisible gateway exhibits an ultra-low effective thermal conductivity below $10^{-3}$ W·m$^{-1}$·K$^{-1}$ (effectively approaching zero), attaining air-equivalent ultrahigh thermal insulation performance, and an ultra-high effective electrical conductivity of up to $2.8\times10^{7}$ S·m$^{-1}$, realizing metal-level high electrical conductivity characteristics. In contrast to conventional low-thermal-conductivity and electrically conductive materials that rely on chemical synthesis and require harsh conditions such as high temperatures, specific crystallographic orientations, or nanostructuring, our work abandons the material modification paradigm and breaks the electrothermal property coupling constraint of natural materials through macroscopic structural regulation. It provides a structural design paradigm and promising engineering application prospects for scenarios requiring efficient electrical conduction with minimal thermal crosstalk, such as on-chip interconnects and wearable electronic devices.

## 1.Introduction

The Wiedemann-Franz law states that the ratio of a material's thermal conductivity to the product of its electrical conductivity and temperature is a constant called Lorenz number, $L = \kappa/\sigma T$ [1-3]. Accordingly, good electrical conductors are often also good thermal conductors (e.g., metals), while thermally insulating materials are usually also electrical insulators (e.g., air). Therefore, conventional materials cannot simultaneously achieve metallic-level electrical conductivity and air-level thermal insulation. However, if new materials with high electrical conductivity and thermal insulation are realized, they hold significant application value in many fields. For example, supplying power to various components in a chip requires highly conductive wiring. If these wires also possess excellent thermal insulation properties, they minimize thermal interference between components while maintaining electrical connections [4]. For wearable electronics, core materials with high electrical conductivity but low thermal conductivity would ensure efficient signal and power transmission while preventing internal heat from reaching the skin, thus enhancing comfort [5]. Furthermore, materials with high electrical conductivity exhibit excellent electromagnetic shielding performance against high-frequency electromagnetic waves [6]. Therefore, materials that are both highly electrically conductive and thermally insulating possess dual characteristics of electromagnetic shielding and thermal insulation against high-frequency electromagnetic waves, promising application value in fields such as electromagnetic device protection and electromagnetic-thermal shielding/cloaking [7-9].

Currently, although some studies have been conducted on new materials that are electrically conductive and exhibit low thermal conductivity, research on these newly synthesized materials primarily relies on chemical synthesis methods [10-19]. For instance, SnSe single crystals demonstrate electrical and thermal conductivities of ~$10^4$ S/m and 0.23–0.34 W/(m·K) along the b-axis at 300–973 K [10], while $VO_2$ nanobeams reach ~$8.0\times10^5$ S/m with 0.72 W/(m·K) at 240–340 K [11]. Chlorine-doped and lead-alloyed n-type SnSe crystals achieve ~$10^5$ S/m and 0.24 W/(m·K) at 300–800 K [12]. Nickel nanoparticles via thermal decomposition exhibit high electrical conductivity of ~$3.57\times10^6$ S/m with 0.9 W/(m·K) at 300 K [13]. Materials such as C/SiCON nonwovens and graphene nanosheet/carbon aerogels show low thermal conductivities of 0.010–0.032 W/(m·K) and 0.027 W/(m·K) with moderate electrical conductivities of 420 S/m and 225 S/m at room temperature, respectively [14,15]. Other systems, including hole-doped SnSe, n-type SnSe, ZrCoBi-based half-Heuslers, and $Cu_{0.01}Bi_2Te_{2.7}Se_{0.3}$-based thermoelectrics, also exhibit balanced electrical conductivities ($10^4$–$10^5$ S/m) and thermal conductivities (1.2–2.2 W/(m·K)) across various temperature ranges [16-19]. However, these methods can only achieve relatively low thermal conductivity (e.g., 0.2–1 W/(m·K)) at room temperature, which is far from the level of air-level thermal insulation (e.g., 0.001 W/(m·K)). Moreover, the electrical conductivity is typically only on the order of thousands of S/m (while the highest conductivity can reach $10^6$ S/m, it is accompanied by relatively high thermal conductivity [13]), making it difficult to achieve metallic-level electrical conductivity. Therefore, there remains a lack of new materials that can simultaneously achieve strong electrical conductivity comparable to metals (e.g., close to copper's electrical conductivity of $5.7\times10^7$ S/m) and strong thermal insulation comparable to air (e.g., close to air's thermal conductivity of 0.024 W/(m·K)).

In recent years, thermal metamaterials and metasurfaces—a class of artificial structural materials designed for thermal regulation [20-22]—have been widely applied in areas such as thermal invisibility cloaks [23-26], thermal concentrators [27-30], thermal illusion devices [31-35], Janus thermal structures [36-39], spatiotemporal thermal devices [40-43], nonreciprocal heat metadevices [44-46], topological state in diffusion systems [47-49] etc. Unlike chemically synthesized materials, whose properties depend on chemical composition, the characteristics of metamaterials and metasurfaces arise from their engineered unit structures [50-53]. This allows them to overcome the limitations of natural materials in thermal conductivity, enabling extraordinary thermal control capabilities—such as tailoring thermal conductivity from near-zero to extremely high values [54], achieving negative thermal conductivity [55], and exhibiting strong anisotropic thermal conduction [56, 57]—that are difficult to realize with conventional chemical synthesis. These advances offer new pathways for heat flow manipulation. Moreover, the field of metamaterials and metasurfaces has evolved rapidly in recent years, expanding from single-physics metamaterials/metasurfaces, which operate within one physical field, to multi-physics metamaterials/metasurfaces. These are composed of the same structural units yet can simultaneously control two or more physical fields [7-9].

To address the long-standing challenge of developing materials that simultaneously achieve metal-level electrical conductivity and air-equivalent thermal insulation, this study draws inspiration from electromagnetic invisible gateways [58-61] and scattering-shifting devices[62-65]. Instead of first designing a thermal invisible gateway with negative thermal conductivity via transformation thermodynamics and spatial folding transformation, we directly adopt the principle of energy conservation to design active thermal metasurfaces (ATMSs). These ATMSs can equivalently realize the functionality of negative thermal conductivity materials required for constructing the thermal invisible gateway [55, 66, 67]. Specifically, the proposed thermal invisible gateway, built on the energy-conservation-derived ATMSs, retains the high electrical conductivity of the metallic substrate while effectively blocking external heat flux, thus realizing the dual property of metal-like electrical conductivity and air-like thermal insulation. Both numerical simulations and experimental results verify the excellent performance of the designed thermal invisible gateway in achieving concurrent high electrical conductivity and superior thermal insulation. Unlike traditional chemically synthesized materials, the proposed structure—integrating an energy-conservation-based thermal invisible gateway into a metal bridge-like substrate—can simultaneously achieve metal-like high electrical conductivity and air-comparable ultrahigh thermal insulation at room temperature.

The thermal invisible gateway is shown in Fig. 1a. It is a symmetric dumbbell-shaped copper bridge structure in air, with carefully designed ATMSs placed between the symmetric square copper plates (colored purple) connecting the left and right sides. When heat flux and electric current are incident simultaneously from left to right, the entire structure exhibits strong electrical conductivity identical to that of the copper substrate—since the copper substrate has high electrical conductivity and the thermal invisible gateway composed of ATMSs has no effect on electric current. Meanwhile, the thermal invisible gateway effectively blocks heat flux (heat flux can barely pass through the bridge structure with ATMSs in the middle), endowing the entire structure with thermal insulation performance close to that of air. In contrast, for the standalone dumbbell-shaped copper bridge structure in air without ATMSs in Fig. 1b, when heat flux and electric current are incident simultaneously from left to right, both can pass through the central bridge structure, meaning the structure simultaneously possesses high electrical conductivity and high thermal conductivity. For the two copper plates with both the central bridge and ATMSs removed in Fig. 1c, when heat flux and electric current are incident simultaneously from left to right, neither can cross the central air region to reach the right side, corresponding to a structure with both low electrical conductivity and low thermal conductivity.

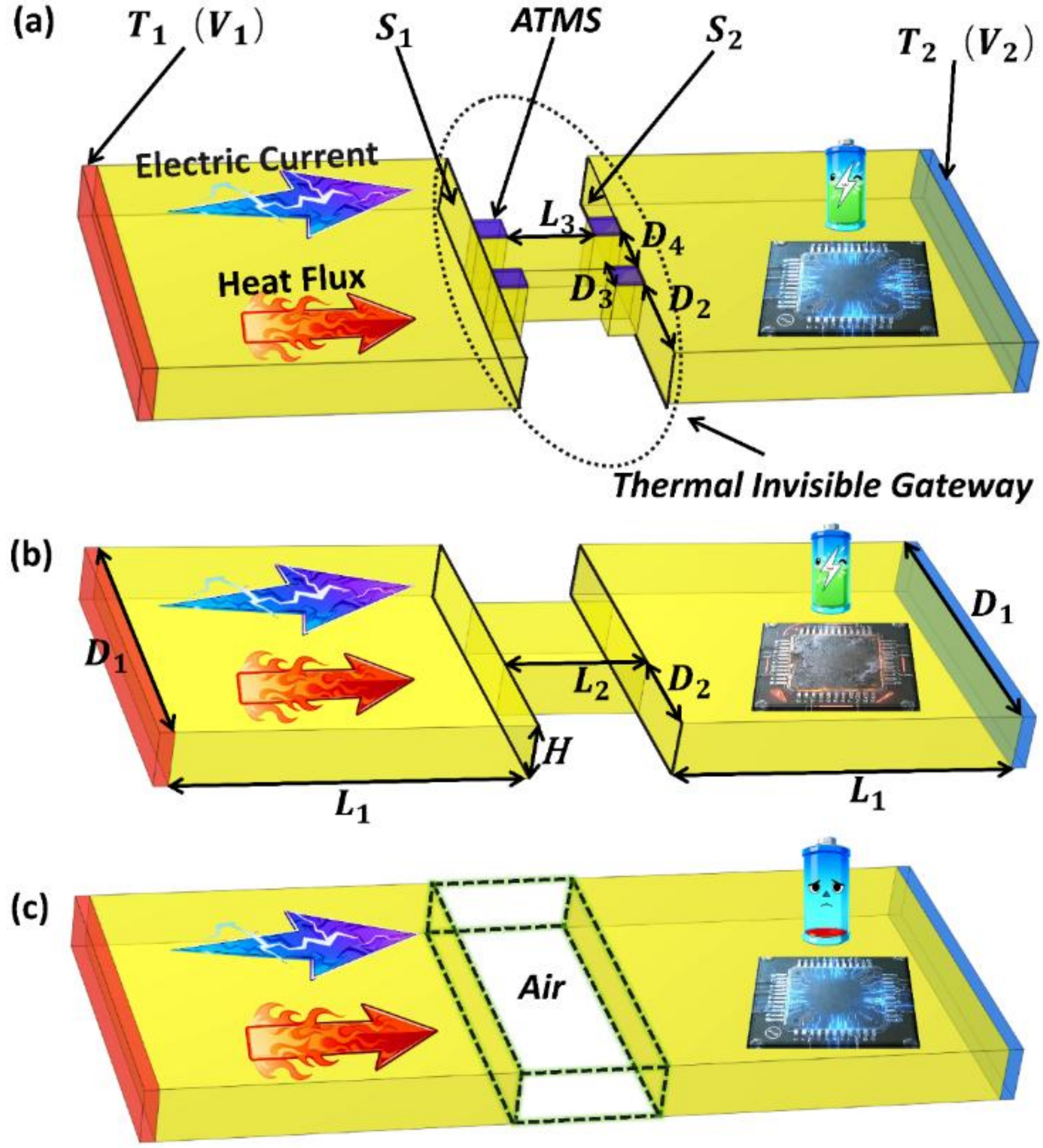

**Fig. 1** (a)–(c) Schematic diagrams of thermoelectric transport behaviors for different structures when heat flux (red arrows) and electric current (blue arrows) propagate from left to right: (a) a symmetric dumbbell-shaped copper bridge structure with a thermal invisible gateway realized by ATMSs (purple squares), where electric current can pass through the central copper bridge to ensure power supply to the right side while heat flux is blocked to prevent excessive temperature rise; (b) a symmetric dumbbell-shaped copper bridge structure without a thermal invisible gateway, where both electric current and heat flux can pass through the central copper bridge, enabling sufficient power supply but also causing an excessively high operating temperature on the right side; (c) a structure with the central bridge removed (two copper plates separated by an air gap) without a thermal invisible gateway, where neither electric current nor heat flux can pass through the central air region, resulting in a normal temperature on the right side but a failure to provide electrical power. Yellow regions represent highly thermally conductive and electrically conductive materials (e.g., metals), white regions represent fully thermally insulating and electrically insulating materials (e.g., air).

## 2. Theoretical Design and Mechanism

The thermal invisible gateway is fabricated on a symmetric dumbbell-shaped copper bridge substrate, as shown in Fig. 1(b). The substrate is placed entirely in air and is prepared by a one-step symmetric cutting of a complete copper plate: from a complete copper plate of total width $D_1$, thickness $H$, and intrinsic thermal conductivity $\kappa$=400 W/(m·K), rectangular regions of length $L_2$, width $D_2$, and thickness $H$ are symmetrically cut from the top and bottom surfaces and filled with air, forming a basic dumbbell-shaped copper bridge. The length of the square copper plates on both sides of the structure is $L_1$; the left side is a constant-temperature heat source $T_1$ (high-potential terminal $V_1$ ), and the right side is a constant-temperature cold source $T_2$ (low-potential terminal $V_2$). The thermal invisible gateway is fabricated by integrating ATMSs onto this substrate through a secondary processing step: inside the bridge region of the basic dumbbell-shaped copper bridge, rectangular regions of length $L_3$, width $D_3$, and thickness $H$ are further symmetrically cut from the top and bottom surfaces and filled with air, forming a stepped boss structure and a continuous conductive channel of width $D_4$ in the bridge region, thus completing the fabrication of the copper matrix structure. Four groups of ATMSs are symmetrically arranged on the bosses of the bridge region: two groups are located at the cross-section $S_1$ near the high-temperature side, and the other two groups are located at the cross-section $S_2$ near the low-temperature side, serving for heat supply and heat absorption, respectively, finally forming the complete thermal invisible gateway structure shown in Fig. 1(a). The core design goal of the structure is to completely block the heat flux in the bridge region while maintaining the continuous conductive path of the copper substrate, thereby achieving ideal thermal isolation between the left high-temperature region and the right low-temperature region.

When the ATMSs operate normally and achieve the ideal thermal invisibility effect, the temperature at the inlet cross-section $S_1$ of the bridge region remains equal to the temperature $T_1$ of the left high-temperature heat source, and the temperature at the outlet cross-section $S_2$ of the bridge region remains equal to the temperature $T_2$ of the right low-temperature heat source. This ideal isothermal boundary condition is the core steady-state premise for the subsequent design of heat flux regulation and power matching. Based on Fourier's law of heat conduction and the law of energy conservation, we carry out the theoretical design, heat flux derivation, and power matching of the thermal invisible gateway. Under the above premise, in the heat conduction process, the heat flux is transmitted along the axial direction from cross-section $S_1$ to cross-section $S_2$ through the bridge region. This path can be uniformly characterized by two series regions, and the total thermal resistance $R_{\mathrm{th}}$ between $S_1$ and $S_2$ is given by:

$$R_{th} = \frac{L_2 - L_3}{\kappa(D_4 + 2D_3)H} + \frac{L_3}{\kappa D_4 H}, \tag{1}$$

where the first term on the right-hand side corresponds to the thermal resistance of the frustum transition section of the bridge region, with a heat flux transmission length of $L_2$-$L_3$ and an effective heat transfer cross-sectional area of $(D_4+2D_3)H$ ; the second term corresponds to the thermal resistance of the middle isolation section of the bridge region, with a heat flux transmission length of $L_3$ and an effective heat transfer cross-sectional area of $D_4H$. Based on the above steady-state temperature boundary condition, when the ATMSs achieve the ideal thermal invisibility effect, the cross-sections $S_1$ and $S_2$ are maintained isothermal with the high-temperature heat source $T_1$ and the low-temperature heat source $T_2$, respectively, so the temperature difference between the two cross-sections is exactly $T_1$-$T_2$. According to Fourier's law of heat conduction and the Ohm's law analogy for heat conduction, the total heat flux $Q$ transmitted axially from $S_1$ to $S_2$ through the bridge region under steady-state conditions is:

$$Q = \frac{T_1 - T_2}{R_{th}} = \frac{\kappa H(T_1 - T_2)(D_4 + 2D_3)D_4}{(L_2 - L_3)D_4 + L_3(D_4 + 2D_3)}. \tag{2}$$

This equation gives the magnitude of the heat flux $Q$ transmitted from cross-section $S_1$ to cross-section $S_2$ when the ideal thermal invisibility effect is achieved.

According to the law of energy conservation, when the thermal invisible gateway achieves the ideal thermal isolation effect, the left high temperature heat source $T_1$ cannot release heat to the bridge region, and the right low temperature heat source $T_2$ cannot receive heat from the bridge region. Therefore, the ATMSs arranged at $S_1$ replace the left heat source and supply an amount of heat $Q$ to the bridge region; meanwhile, the ATMSs arranged at $S_2$ absorb this amount of heat to prevent it from entering the right low temperature region. Then the total output power $P_1$ of the ATMSs at $S_1$ and the total output power $P_2$ of the ATMSs at $S_2$ satisfy:

$$\begin{cases} P_1 = Q \\ P_2 = -Q \end{cases}. \tag{3}$$

When this power matching condition is satisfied, the heat transmitted to cross-section $S_2$ will be completely absorbed and cannot enter the right low-temperature region, and the preset temperature $T_2$ can be stably maintained at $S_2$. The above regulation process satisfies strict energy conservation: the heat supplied by the ATMSs at $S_1$ is equal in magnitude to the heat absorbed by the ATMSs at $S_2$, with no additional energy input and loss in the system. The directional heat supply and absorption inside the bridge region are realized only through ATMSs, which replaces natural heat conduction and achieves thermal isolation between the two sides.

Notably, it can be seen from the heat flux and power matching formulas that the regulation power of ATMSs is only determined by the temperature difference $T_1$- $T_2$ between the two sides of the structure, and is independent of the specific temperature values of $T_1$ and $T_2$. No matter how the temperature of the heat sources on both sides changes, as long as the temperature difference between them remains constant, the required regulation power of ATMSs does not need to be adjusted, which endows the structure with extremely strong working condition adaptability. In practical applications, the heat blocking effect can be guaranteed as long as the total power meets the above matching conditions, and the change in the number of units has no significant impact. The structure has excellent material and scenario adaptability: the copper matrix can be replaced by most conductive materials, and the air filling layer can be replaced by most insulating materials.

By combining macroscopic structural design with ATMSs, the thermal invisible gateway breaks the constraint of intrinsic coupling between electrical and thermal properties of natural materials, and synergistically achieves metallic-level high electrical conductivity and air-level ultrahigh thermal insulation performance in the overall structure. The physical essence originates from the fundamental difference in the transmission mechanism of heat and electricity in metals, as well as the differentiated regulation of their transmission paths. In the copper substrate, both electric current and heat flux are carried by free electrons, but their transmission laws and regulation methods are essentially different: the transmission of electric current relies on the directional drift motion of free electrons under an applied electric field, and the core premise for continuous transmission is a complete conductive path. Only non-through local cutting is performed on the copper substrate, and the copper matrix with width $D_4$ in the bridge region always maintains a continuous and complete conductive path. ATMSs only regulate the thermal field, which will not damage the conductive path of the copper substrate or interfere with the directional drift process of free electrons. Therefore, the structure always retains the metallic-level high electrical conductivity of the pure copper substrate, and the electric current can be smoothly transmitted from the left $V_1$ terminal to the right $V_2$ terminal. In contrast, the transmission of heat flux relies on the random thermal motion of free electrons, and the spontaneous diffusion of energy is realized through collisions between electrons, which strictly follows Fourier's law of heat conduction. Relying on the precise power matching of ATMSs, heat is supplied to the bridge region through positive power output at cross-section $S_1$, and the heat transmitted to the cross-section is completely absorbed through negative power output at cross-section $S_2$. This replaces natural heat conduction by means of internal energy conservation, completely blocks the heat transfer from the left high-temperature region to the right low-temperature region, makes the effective thermal conductivity of the bridge region approach zero, and finally achieves ultrahigh thermal insulation performance equivalent to that of air. By limiting the concentrated transmission path of heat flux through structural optimization, realizing directional heat supply and absorption through precise power matching of ATMSs, and completely retaining the continuous conductive path of the metal substrate at the same time, this work provides a new structural design paradigm for engineering scenarios that require both efficient electrical conduction and extreme thermal insulation.

# 3.Numerical Simulation and Parameter Optimization

To verify the function of the thermal invisible gateway in achieving high electrical conductivity and thermal insulation in the copper bridge structure, numerical simulations are performed via the finite element method using COMSOL Multiphysics 5.6. By incorporating the solid heat transfer module and electric current module, we simulate the transport processes of heat flux and electric current (the detailed numerical settings are given in **Supplementary Document 1**). To verify that the structure designed in Figure 1a possesses air-equivalent thermal insulation and metal-level electrical conductivity, we primarily simulate the steady-state temperature field and steady-state electric potential field distributions when heat flux and electric current are incident on the structure from left to right. For comparison, we also simulate the corresponding distributions for the standalone copper bridge (Figure 1b) and the air-isolated structure without a copper bridge (Figure 1c). As shown in the upper panel of Figure 2a, the steady-state temperature field distribution of the copper bridge structure integrated with the thermal invisible gateway (structure in Figure 1a), after reaching thermal equilibrium, is consistent with that of the air-isolated structure (structure in Figure 1c, simulated in the upper panel of Figure 2c). This is significantly different from the scenario simulated in the upper panel of Figure 2b, where heat flux flows from left to right through the copper bridge (structure in Figure 1b). This result verifies that the copper bridge structure with the thermal invisible gateway achieves thermal insulation performance comparable to that of air. Meanwhile, the lower panel of Fig. 2a presents the simulated steady-state electric potential field distribution of the copper bridge structure with the thermal invisible gateway (the structure in Fig. 1a), which is identical to that of the standalone copper bridge structure shown in the lower panel of Fig. 2b, where electric current can smoothly pass through the copper bridge in both structures. This distribution is significantly different from that of the air-isolated structure displayed in the lower panel of Fig. 2c, in which the electric current is blocked by the intermediate air gap. This result demonstrates that the copper bridge structure integrated with the thermal invisible gateway can maintain metal-level electrical conductivity consistent with that of the pure copper bridge.

To quantitatively describe the heat transfer and electrical conduction performance of the copper bridge structure integrated with the thermal invisible gateway, and to further explore the relationship between the structure's geometric parameters and its thermal insulation and electrical conduction characteristics, we introduce the equivalent thermal conductivity ($\kappa$) and equivalent electrical conductivity ($\sigma$) of the whole structure, respectively:

$$
\begin{cases}
\kappa = \dfrac{Q \cdot L}{\Delta T \cdot S}, \\
\sigma = \dfrac{I \cdot L}{\Delta U \cdot S},
\end{cases}
\tag{4}
$$

where $Q$ is the total heat quantity flowing from the high-temperature source to the low-temperature source per unit time. $I$ is the total electric current intensity. The distance between the high-temperature (high-potential) source and the low-temperature (low-potential) source is equal to the total length of the structure $L=2L_1+L_2$. $\Delta T=T_1-T_2$ and $\Delta U=V_1-V_2$ are the temperature difference and voltage difference between the two boundary planes loaded with thermal and electrical sources, respectively. $S$ is the cross-sectional area of this symmetric dumbbell-shaped copper bridge structure at the aforementioned two boundary planes ($S=S_1=S_2$). Based on the above definitions of the equivalent parameters, we further investigate how the equivalent thermal conductivity and equivalent electrical conductivity vary with the core geometric parameters of the thermal invisible gateway.

During the two-dimensional parameter scanning, the basic geometric parameters of the copper bridge structure ($L_1$, $D_1$, $D_3$) are fixed, while two core geometric parameters of the proposed thermal invisible gateway—its length $L_2$ and its main width $D_4$ (marked in Figure 1a)—are individually adjusted in a gradient manner, with the theoretical target power of ATMSs for achieving zero equivalent thermal conductivity calculated synchronously via the previously derived Equation (2) to serve as a reference for simulation tuning. Figures 2(d)–(f) present the scanning results of $L_2$, $D_4$, and total ATMSs output power, respectively. For the gateway length $L_2$ scanned from 0.25 dm to 0.65 dm (Figure 2(d)), both the theoretical and simulated power required for zero thermal conductivity show a consistent downward trend with increasing $L_2$, as the extended gateway length lengthens the axial heat transfer path, linearly increases the total thermal resistance of the bridge region, reduces the temperature-driven spontaneous heat flux, and thus lowers the required compensation power synchronously. For the gateway's main width $D_4$ scanned from 0.10 dm to 0.45 dm (Figure 2(e)), the required power for zero thermal conductivity increases monotonically with $D_4$, since a wider gateway main width expands the effective heat transfer cross-section, reduces thermal resistance, enhances spontaneous heat transfer, and therefore demands higher compensation power. Notably, despite the slight deviation between theoretical and simulated values, the zero equivalent thermal conductivity state can always be achieved by fine-tuning the actual ATMSs power based on the theoretical reference for structures with different geometric parameters, verifying the strong robustness and geometric adaptability of the proposed thermal regulation strategy. Figure 2(f) presents the variation of equivalent thermal conductivity $\kappa$ and electrical conductivity $\sigma$ with total ATMSs power: the equivalent $\sigma$ remains nearly constant at a high value close to the intrinsic conductivity of copper, as ATMSs only regulate the thermal field without damaging the continuous conductive path of the copper substrate, while $\kappa$ decreases linearly with increasing ATMSs power, showing positive $\kappa$ at lower power and negative $\kappa$ at higher power; the intersection of the $\kappa$ curve with the zero thermal conductivity axis corresponds to the actual power required for ideal thermal isolation in the simulation, and the red star marks the theoretical zero-thermal-conductivity power from Equation (2), which is higher than the simulated value, consistent with the deviation trend in Figures 2(d) and 2(e).

The core cause of this deviation lies in the essential difference between the one-dimensional ideal theoretical model and the two-dimensional simulation setup. The theoretical calculation assumes that all spontaneous heat flux generated at the high-temperature side must be fully transmitted to the low-temperature side through the axial channel of the copper bridge, thus requiring the ATMSs to output full power to completely cancel this heat flux. The theoretical value in Equation (2) is derived based on this ideal full-compensation assumption, and the thermal effect of the surrounding air is completely ignored in the calculation. However, in the actual two-dimensional simulation, the thermophysical parameters of air are fully incorporated, and the surrounding air effectively "dissipates" part of the heat flux: a portion of the spontaneous heat flux is directly released into the ambient environment through the copper-air interface at the side of the copper bridge, without passing through the axial channel or requiring compensation from the ATMSs. Meanwhile, the inherent low thermal conductivity of air provides additional thermal resistance at the edge of the copper bridge, further reducing the net heat flux that needs to be transmitted through the axial channel. Therefore, the total heat flux to be compensated by the ATMSs in the simulation is naturally smaller than the full value obtained from the theoretical calculation. In addition, the theoretical model assumes an ideal continuous surface heat source, while the simulation adopts four discrete point heat sources (one per boss) precisely aligned with the axial heat flux path, which delivers more targeted heat compensation and achieves complete cancellation of the axial heat flux with lower output power, which is also an important reason for the theoretical value being higher than the actual simulation value.

It should be emphasized that for the thermal invisible gateway proposed in this work, structures with arbitrary geometric parameters can achieve the ideal state of zero equivalent thermal conductivity by fine-tuning the total output power $Q$ of ATMSs, without being limited to specific geometric dimensions. The structural parameters finally determined for the experiment are: $D_2$=0.40 dm, $L_2$=0.60 dm, $D_3$=0.10 dm, $L_1$=1.50 dm, $D_1$=1.20 dm, $L_3$=0.40 dm, $D_4$=0.20 dm. The selection of this set of parameters is comprehensively based on the convenience of laser processing of the copper substrate, and the 0.20 dm length reserved at both ends of the copper sheet for the installation of high-temperature and low-temperature source devices in the experiment, which ensures the feasibility of sample preparation and experimental measurement.

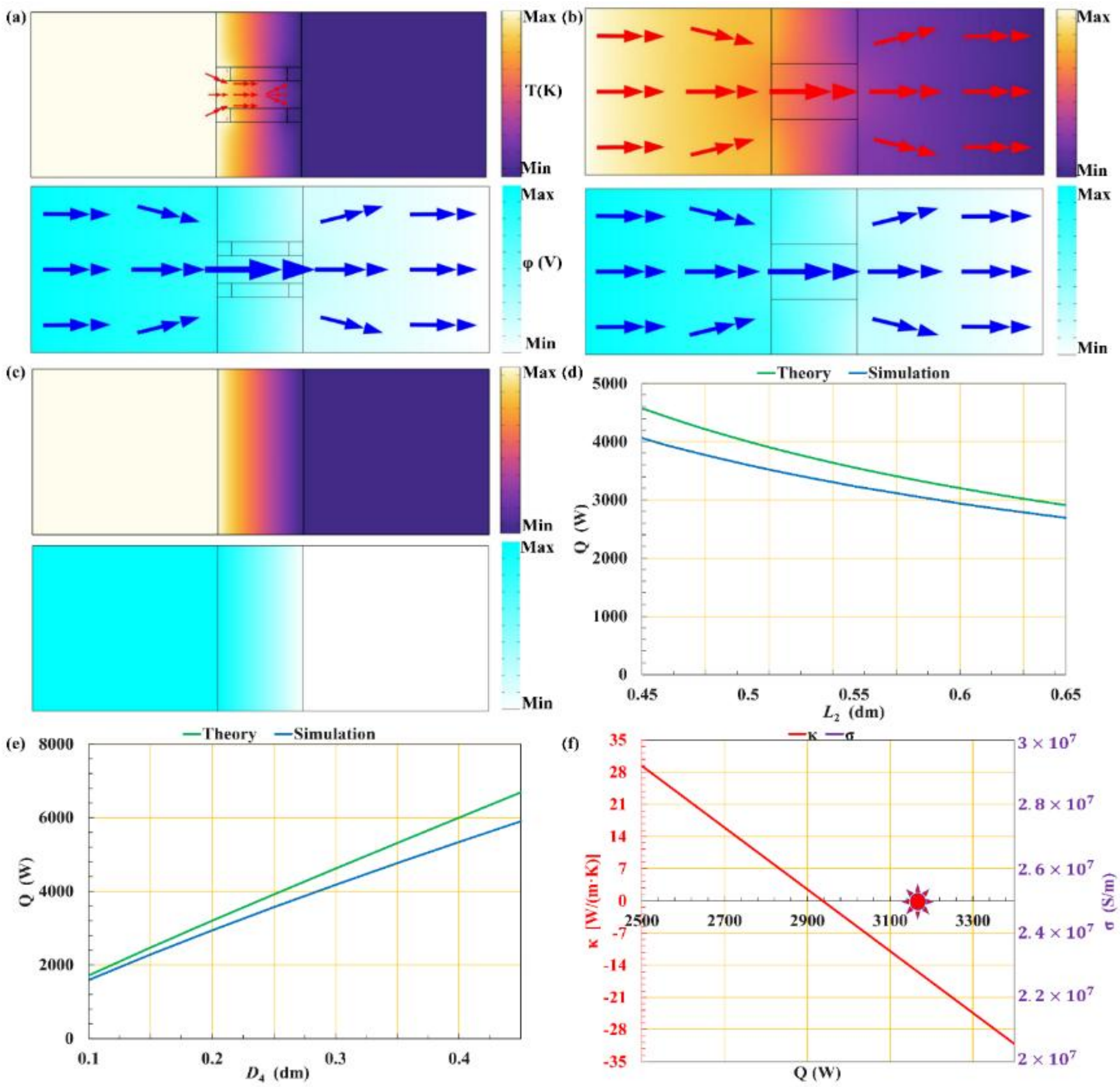

**Fig. 2** (a)-(c) Two-dimensional simulated nephograms of temperature field and electric potential field for different structures at thermal/electrical equilibrium when heat flux and electric current propagate from left to right. Specifically, (a) corresponds to the simulated distribution of temperature and electric potential for the structure with high electrical conductivity and thermal insulation (integrated with thermal invisible gateway) at equilibrium, (b) for the structure with simultaneous electrical and thermal conductivity (pure copper bridge without thermal invisible gateway) at equilibrium, and (c) for the structure with neither electrical nor thermal conductivity (two copper plates separated by air without copper bridge) at equilibrium. The geometric parameters adopted for all structures in the simulation are: $L_1$=0.13 m, $D_1$=0.12 m, $L_2$=0.06 m, $D_2$=0.04 m, $L_3$=0.04 m, $D_3$=0.01 m, $D_4$=0.02 m. (d) and (e) show the comparison between the theoretical heat flux $Q$ (green curves, derived from Eq. (2) at zero equivalent thermal conductivity) and the simulated $Q$ (blue curves, under the same zero thermal conductivity condition) with the variation of bridge interface length $L_2$ and conductive channel width $D_4$, respectively; (f) presents the variation curves of equivalent thermal conductivity $\kappa$ (red curve) and equivalent electrical conductivity $\sigma$ (purple curve) with the total power of ATMSs $Q$, where the red star marks the theoretical $Q$ value corresponding to zero equivalent thermal conductivity, and the remaining geometric parameters are kept consistent with the above initial simulation parameters during each single-parameter scan.

# 4.Experimental Preparation and Measurement

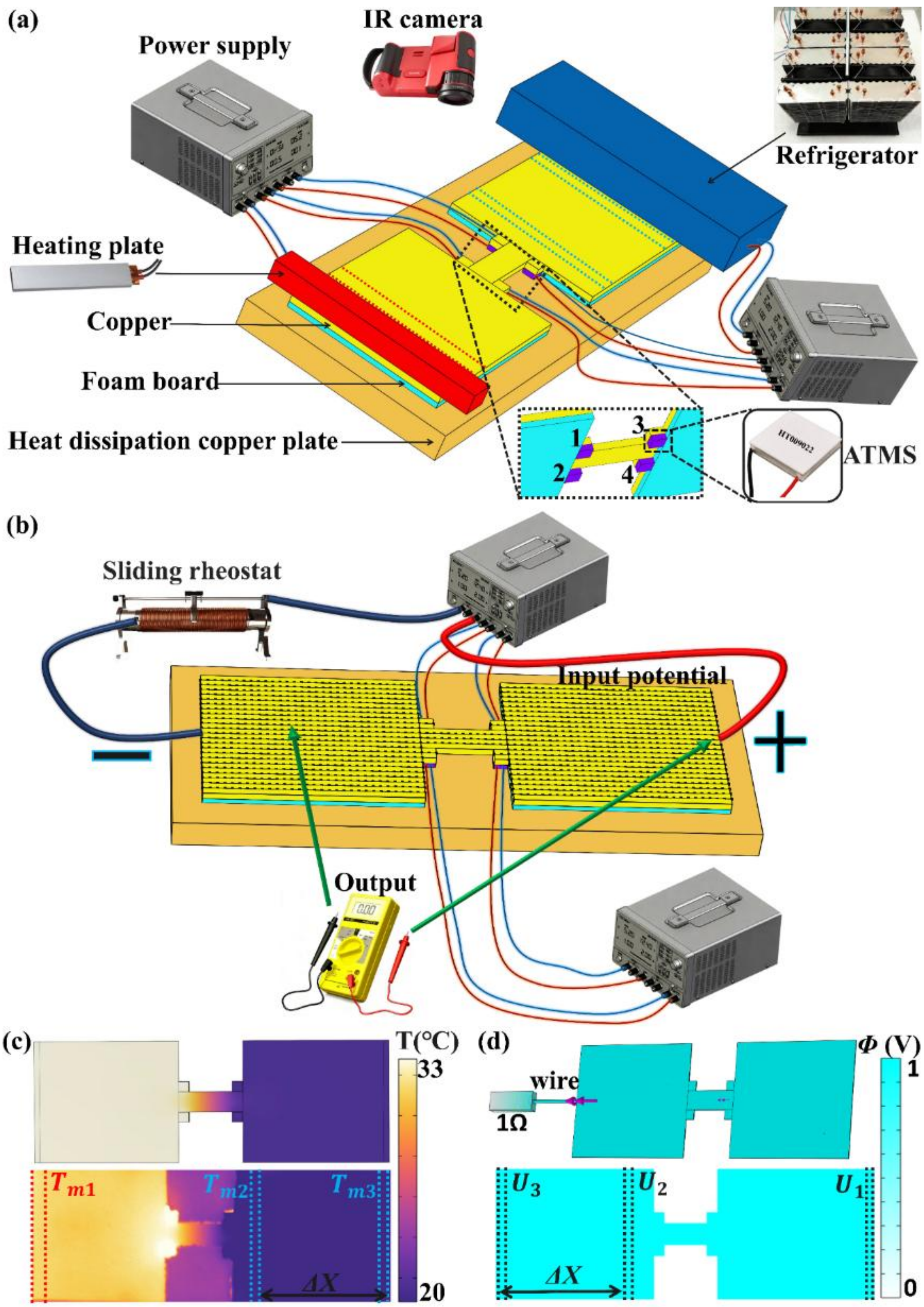

**Fig. 3** Experimental verification of the thermal and electrical performance of the thermal invisible gateway. (a) Schematic of the thermal test setup. The dumbbell-shaped copper substrate shown in Fig. 1(a) is used as the core component, with a heating plate and a cooling device at its two ends. Four ATMSs are mounted on the square bosses at the bridge interface to achieve equivalent negative thermal conductivity. The copper substrate is insulated by expanded polystyrene foam and coupled to a heat-dissipating copper substrate kept at 23 ℃. An infrared thermal imager monitors the surface temperature field, with red and blue dashed boxes indicating the high- and low-temperature measurement regions. (b) Schematic of the electrical test setup. After removing the thermal devices, the copper substrate is connected in series with a 1 Ω rheostat and a 1 V power supply. Potential

measurements are performed with a high-precision multimeter, where one probe is fixed at the high-potential end and the other contacts the black dots (preset measurement points) on the copper plate; "+" and "-" denote the positive and negative poles of the power supply. (c) Temperature field distribution of the copper substrate under thermal steady state. The upper and lower panels present the simulation and experimental results with a unified temperature scale. The three dashed boxes represent the temperature measurement regions, with average temperatures labeled as $T_{m1}$, $T_{m2}$ and $T_{m3}$, respectively, where $T_{m1}$=32.8 °C is the average temperature of the high-temperature side region, $T_{m2}$= $T_{m3}$=20.8 °C are the average temperatures of the two low-temperature side regions. (d) Potential field distribution under electrical steady state. The upper and lower panels show the simulation and experimental results with a unified potential scale. The three black dashed boxes represent the potential measurement regions, with average potentials labeled as $U_1$, $U_2$ and $U_3$, respectively. The distance between the $T_{m2}$ and $T_{m3}$ temperature measurement regions, as well as that between the $U_2$ and $U_3$ potential measurement regions, is uniformly $\Delta X$=0.12 m.

Sample preparation is conducted in accordance with the structural dimensions shown in Fig. 1(a). A dumbbell-shaped copper substrate is first fabricated via laser cutting, with the specific geometric parameters set as $D_2$=0.04 m, $L_2$=0.06 m, $D_3$=0.01 m, $L_1$=0.15 m, $D_1$=0.12 m, $L_3$=0.04 m and thickness $H$=0.001 m to ensure structural machining precision. Subsequently, in a constant temperature room, a thin layer of black paint is uniformly sprayed on the infrared observation surface of the copper plate to improve the contrast of infrared imaging, followed by a 2-hour standing period for complete air-drying of the paint. ATMSs (Model HT009022) in the form of thermoelectric components are mounted on the four square bosses in the bridge interface area on the non-observation surface of the copper plate, respectively. These ATMSs operate based on the Peltier effect and can release or absorb heat by applying forward or reverse electric currents. To eliminate air gaps at the contact interface between ATMSs and the copper plate and ensure heat transfer efficiency, thermal conductive silicone grease VK887 is evenly applied at the contact positions, and then the ATMSs are firmly bonded to the copper plate surface with adhesive tape and solid glue. After that, the remaining areas on this side of the copper plate, except for the bridge interface area and ATMSs, are covered with an expanded polystyrene (EPS) foam board with a thickness of 2.6 mm and a thermal conductivity of 0.05 W/(m·K) and fixed with adhesive tape, which not only achieves thermal insulation protection for non-test areas but also provides mechanical support for the overall structure. Finally, the copper plate assembled with ATMSs is coupled and attached to the heat-dissipating copper substrate via thermal conductive silicone grease with the mounting surface facing downward. During the experiment, the heat-dissipating copper substrate is always maintained at an ambient temperature of approximately 23 °C, which can stably absorb or transfer the heat from the bottom surface of ATMSs and ensure that ATMSs operate in a steady state at all times (**Movie 1**).

After sample preparation, pre-experiment debugging and arrangement are carried out. First, the heating plate and the refrigeration device are closely attached and fixed to the corresponding positions on both sides of the test copper plate, respectively, and independent power supplies are connected to each of them. Then, the four ATMSs are connected to two adjustable DC power supplies with multi-channel output via alligator clip wires to realize independent power supply control for each ATMS unit. To minimize the interference of air convection on the experimental measurement results, the entire experiment is conducted in a closed air-conditioned laboratory, with the indoor ambient temperature stably maintained at approximately 23 °C (**Supplementary Document 2**).

The experimental measurements are divided into two parts: thermal characteristic measurement (see Fig. 3a) and electrical characteristic measurement (see Fig. 3b), with the overall operation process as follows. The prepared sample is placed at the preset experimental position, and the infrared thermal imager (FOTRIC 288) mounted on a tripod is adjusted to ensure its field of view completely covers the detection area on the sample surface. After turning on the infrared thermal imager, the heating plate and refrigeration device connected to independent DC constant current power supplies are activated, and their output currents are adjusted according to the real-time monitoring data of the infrared thermal imager until the temperatures of the high-temperature side and low-temperature side of the copper plate are stably maintained at 32.8 °C and 20.8 °C, respectively, achieving a constant temperature difference of 12 K. Subsequently, independent DC constant current power supplies are connected to the four ATMSs units, and the initial current value of each unit is set according to the theoretical estimation formulas $Q_H=\alpha IT_H+0.5I^2R-\kappa_T\Delta T$ and $Q_L=\alpha IT_L-0.5I^2R-\kappa_T\Delta T$ under the nonlinear approximation condition (**Supplementary Document 3**).

After the system reaches thermal steady state (usually taking 35 seconds), the temperature distribution data on the copper plate surface are recorded by the infrared thermal imager. After recording, all power supplies are turned off to allow the system to cool naturally to the ambient temperature. The measured temperature data are compared with the numerical simulation results, and the supply current of ATMSs is fine-tuned accordingly based on the deviation. After eight rounds of repeated measurements and slight current adjustments, the optimized current value of each ATMS is determined (**Supplementary Document 3**), and the temperature distribution obtained from the actual measurement is consistent with the simulation results at this time. This fine-tuning operation is necessary because the output power of ATMSs has a nonlinear relationship with the supply current and there are minor process tolerances during sample preparation. After applying the optimized current and allowing the system to reach thermal steady state again, the final temperature distribution of the copper plate is captured and is shown in Fig. 3c. After completing the thermal characteristic measurement, the heating plate and refrigeration device attached to both sides of the copper plate are removed, and the installation position and power supply state of ATMSs are kept unchanged. The copper plate is connected in series with a sliding rheostat with a preset resistance of 1 Ω and a power supply with a constant voltage output of 1 V to build an electrical measurement circuit. A high-precision digital multimeter is used for potential measurement: one test lead is fixed at the high-potential end of the copper plate, and the other test lead is in contact with each preset measurement point on the copper plate in turn to record the voltage difference of each point relative to the high-potential end. All measured potential data are sorted and recorded in Microsoft Excel, then imported into COMSOL Multiphysics for subsequent processing, and the potential distribution of the copper plate is finally obtained as shown in Fig. 3d (**Movie 2**).

To further extract the effective thermal conductivity of the structure from the experimental results, this study imports the surface temperature data of the dumbbell-shaped copper plate used in the experiment, collected by an infrared thermal imager, into the Analyzer software for quantitative measurement and analysis. Since the heat injected into the low-temperature source of the copper plate cannot be directly measured, we select two temperature measurement regions on the low-temperature side of the copper plate (the areas marked by the blue dashed boxes in Fig. 3c) for analysis. The temperature field distribution in these regions is uniform, based on which the heat can be calculated via Fourier's law of heat conduction through the temperature difference between the two regions ($T_{m2}$ - $T_{m3}$), the distance between the regions $\Delta X$, and the cross-sectional area of the structure $S_2$; the calculation formula for the heat injected into the low-temperature source of the copper plate is exactly the expression in the square brackets of Equation (7). Combined with the above measured parameters, as well as the core geometric parameters of the structure itself $L_1$ and $L_2$, and the intrinsic thermal conductivity of the copper plate $\kappa_1$=400 W/(m·K), the calculation formula for the overall effective thermal conductivity $\kappa_{eff}$ of the structure is established as follows:

$$\kappa_{eff} = \frac{(2L_1 + L_2)}{(T_{m1} - T_{m3}) \cdot S_2} \cdot \left[ \frac{(T_{m2} - T_{m3}) \cdot \kappa_1 \cdot S_2}{\Delta X} \right] = \frac{(2L_1 + L_2) \cdot (T_{m2} - T_{m3}) \cdot \kappa_1}{(T_{m1} - T_{m3}) \cdot \Delta X}. \quad (5)$$

Through the calculation with the above formula, the overall effective thermal conductivity of the structure is approximately derived as $\kappa_{eff}$=0.

To further extract the effective electrical conductivity of the structure from the experimental results, comprehensive measurement and analysis are performed on the measured data of all potential measurement points. Since the current injected into the low-potential terminal of the copper plate cannot be directly measured, we select two potential measurement regions on the low-potential side of the copper plate (the areas marked by the black dashed boxes in Fig. 3d) for analysis. The potential field distribution in these regions is uniform, based on which the current can be calculated via Ohm's law through the potential difference between the two regions ($U_2$ - $U_3$), the distance between the regions $\Delta X$, and the cross-sectional area of the structure $S_2$; the calculation formula for the current injected into the low-potential terminal of the copper plate is exactly the expression in the square brackets of Equation (8). Combined with the above measured parameters, the inherent parameters of the structure itself, and the intrinsic electrical conductivity of the copper plate $\sigma_1$=5.96×10$^7$ S/m, the calculation formula for the overall effective electrical conductivity $\sigma_{eff}$ of the structure is established as follows:

$$\sigma_{eff} = \frac{(2L_1 + L_2)}{(U_1 - U_3) \cdot S_2} \cdot \left[ \frac{(U_2 - U_3) \cdot \sigma_1 \cdot S_2}{\Delta X} \right] = \frac{(2L_1 + L_2) \cdot (U_2 - U_3) \cdot \sigma_1}{(U_1 - U_3) \cdot \Delta X}. \quad (6)$$

Through the calculation with the above formula, the overall effective electrical conductivity of the structure is derived as approximately $\sigma_{eff}$=2.8×10$^7$ S/m.

# 5. Discussion

We have numerically and experimentally demonstrated a thermal invisible gateway on a copper bridge substrate, integrated with ATMSs that enables metal-level high electrical conductivity and air-equivalent ultrahigh thermal insulation. Existing research on materials with conductive and low-thermal-conductivity characteristics mostly relies on chemical synthesis and structural regulation methods, often accompanied by harsh implementation conditions (see Fig. 4). Systematic quantification and comparison of its key performance metrics and implementation conditions with reported conductive low-thermal-conductivity materials further highlight its superior electrothermal regulation performance, laying a direct foundation for performance and paradigm analysis.

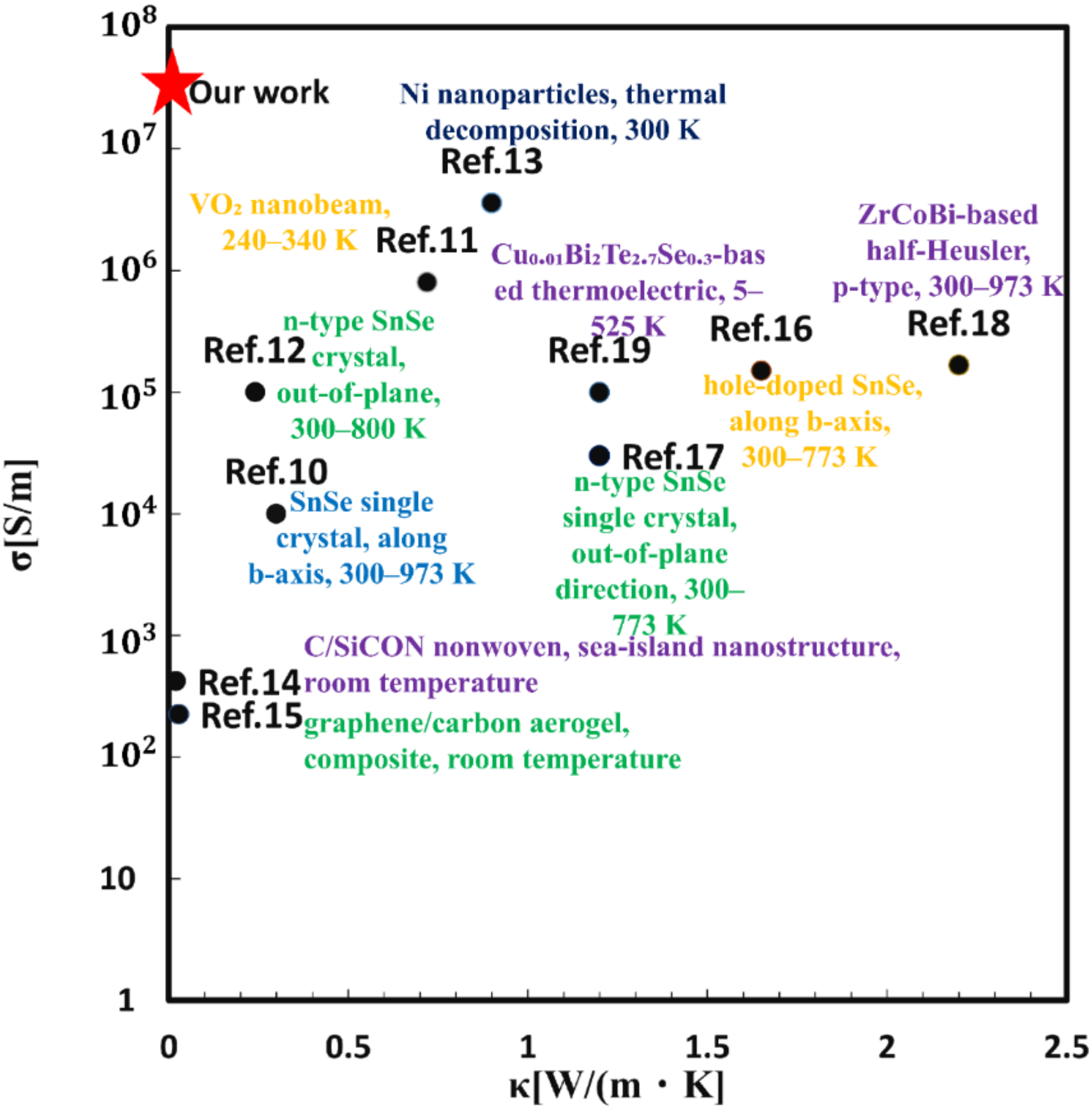

**Fig. 4** Electrothermal performance comparison between the thermal invisible gateway in this work and reported conductive low-thermal-conductivity materials. The red star represents our work, and the other data points correspond to the performance of various materials reported in the references. The y-axis is the electrical conductivity ($\sigma$, S/m, logarithmic scale), and the x-axis is the thermal conductivity ($\kappa$, W/(m·K)).

## 6.Conclusion

In conclusion, we address the longstanding fundamental challenge imposed by the Wiedemann-Franz law: the intrinsic coupling of electrical and thermal conductivity in conventional materials, which prevents the simultaneous realization of metal-level high electrical conductivity and air-equivalent ultrahigh thermal insulation. We resolve this challenge by designing, fabricating, and experimentally validating a thermal invisible gateway based on ATMSs. Guided by the principle of energy conservation and Fourier's law of heat conduction, we theoretically derive the heat flux compensation power requirements for ATMSs, optimize the structural geometric parameters via finite element simulations, and fabricate a macroscopic copper-based prototype that operates stably at room temperature. Experimental measurements demonstrate that our device achieves an ultra-low effective thermal conductivity below $10^{-3}$ $W \cdot m^{-1} \cdot K^{-1}$ (effectively approaching zero) alongside a high effective electrical conductivity of $2.8\times10^{7}$ $S \cdot m^{-1}$, with excellent agreement between simulation and experimental results. Unlike state-of-the-art conductive low-thermal-conductivity materials that rely on chemical synthesis and require harsh operating conditions such as high temperatures, specific crystallographic orientations or nanostructuring, our work realizes the physical decoupling of thermal and electrical transport pathways in a macroscopic metallic substrate through a structural regulation paradigm combined with active heat flux compensation, breaking the inherent electrothermal coupling constraint of the Wiedemann-Franz law on traditional material systems. This thermal invisible gateway features excellent material and scenario adaptability, and its unprecedented synergy of metal-level electrical conductivity and air-equivalent thermal insulation establishes a new design framework for high-density electronic packaging, chip thermal management, flexible wearable electronics and electromagnetic-thermal dual shielding, while also providing critical theoretical insights and technical routes for the development of active thermal metamaterials and multi-physics field regulation devices.

## Acknowledgments

This work is supported by the National Natural Science Foundation of China (Nos. 12274317, and 12374277), San Jin Talent Support Program—Shanxi Provincial Youth Top-notch Talent Project, the Natural Science Foundation of Shanxi Province (202303021211054), and Shanxi Province Higher Education Institutions Young Faculty Research and Innovation Support Program (2025Q006).